\documentclass[english,pre,twocolumn]{revtex4}
\usepackage[T1]{fontenc}
\usepackage[latin9]{inputenc}
\usepackage{amsmath}
\usepackage{amssymb}
\usepackage{graphicx}
\usepackage{esint}

\makeatletter

\@ifundefined{textcolor}{}
{%
 \definecolor{BLACK}{gray}{0}
 \definecolor{WHITE}{gray}{1}
 \definecolor{RED}{rgb}{1,0,0}
 \definecolor{GREEN}{rgb}{0,1,0}
 \definecolor{BLUE}{rgb}{0,0,1}
 \definecolor{CYAN}{cmyk}{1,0,0,0}
 \definecolor{MAGENTA}{cmyk}{0,1,0,0}
 \definecolor{YELLOW}{cmyk}{0,0,1,0}
}

\makeatother

\usepackage{babel}
\begin{document}

\title{Exact results for a noise-induced bistable system}

\author{Bahram Houchmandzadeh $^{1,2}$ and Marcel Vallade$^{1,2}$.}

\affiliation{(1) Univ. Grenoble Alpes, LIPHY, F-38000 Grenoble, France.\\
 (2) CNRS, LIPHY, F-38000 Grenoble, France. }

\begin{abstract}
A stochastic system where bistability is \emph{caused} by noise\emph{
}has been recently investigated by Biancalani et \emph{al.} (PRL 112:038101,
2014). They have computed the mean switching time for such a system
using a continuous Fokker-Planck equation derived from the Taylor
expansion of the Master equation to estimate the parameter of such
a system from experiment. In this article, we provide the exact solution
for the full discrete system without resorting to continuous approximation
and obtain the expression for the mean switching time. We further
extend this investigation by solving exactly the Master equation and
obtaining the expression of other quantities of interests such as
the dynamics of the moments and the equilibrium time. 
\end{abstract}
\maketitle

\section{Introduction.}

In some stochastic systems, noise can have counter intuitive effects
and the behavior of the system be markedly different from its deterministic,
mean field approximations. In some oscillatory gene networks, the
regular oscillations are caused by noise and cease in their absence
\cite{Vilar2002}. In population genetics, the noise term can explain
the emergence of less fit ``altruistic'' individuals \cite{Houchmandzadeh2012a}.
In ecology, the spatial aggregation of individuals can be caused by
noise \cite{Korolev2010,Houchmandzadeh2008} ; a similar explanation
lies behind neutron clustering in nuclear reactors \cite{Dumonteil2014}. 

The general theory of noise induced transition in non-equilibrium
systems has been extensively investigated by Horsthemke and Lefeve
\cite{Horsthemke}. In the context of chemical equations and specifically
genetic regulatory networks, there has been an intense investigation
of systems where \emph{bistability} is caused by noise and is absent
from the deterministic formulation of kinetic rate equations. Samoilov
et al. \cite{Samoilov2005} have considered the enzymatic futile cycle
reaction and have shown that addition of noise can cause bistability
and dynamic switching in the concentration of the substrate. Artyomov
et al. \cite{Artyomov2007} have considered a simple model of T cells
response and have shown again that in the presence of noise, the steady
state distribution can become bi-modal. Qian et al. \cite{Qian2009}
and Thomas et al. \cite{Thomas2014}, using different approaches,
have derived a general framework to elicit the role of fluctuation
time scales separation in the appearance of noise induced bistability.
In an elegant experiment, To and Maheshri \cite{To2010} have investigated
a synthetic transcriptional feedback loop and have demonstrated the
bimodality of the response \emph{without} cooperative binding of the
transcription factor, a usual hypothesis to explain bistability of
genetic switches. 

Recently, Biancalani et \emph{al.} \cite{Biancalani2014}  investigated
another stochastic system where bistability is caused by noise: in
this system, individuals (or molecules) can be in one of the two configurations
$A$ and $B$ and can switch from one to the other according to the
following transition rates:
\begin{eqnarray}
W^{-}(n) & = & W(n\rightarrow n-1)=\left(r(N-n)+\epsilon\right)n\label{eq:Wm}\\
W^{+}(n) & = & W(n\rightarrow n+1)=\left(rn+\epsilon\right)(N-n)\label{eq:Wp}
\end{eqnarray}
where $n$ is the number of individuals in configuration $A$ and
$N$ is the total number of individuals. In the following, $n$ is
used to characterize the state of the stochastic system at a given
time. The rate $r$ characterizes the two body interactions 
\[
X_{i}+X_{j}\xrightarrow{r}2X_{i}\,\,\, i=A,B;\,\, j=B,A
\]
while the rate $\epsilon$ characterizes spontaneous switching of
an individual from one configuration to the other:
\[
X_{i}\xrightarrow{\epsilon}X_{j}
\]
Without loss of generality, we will set $r=1$ in the following. This
is achieved by scaling both time and $\epsilon$ by the factor $r$. 

Such a system can model for example a colony of foraging ants collecting
food from two sources. In population genetics, this is the Moran model
for two competing alleles $A$ and $B$ with bidirectional mutations
\cite{Moran1962}. Such systems were also proposed in the context
of autocatalytic chemical reactions with small number of molecules
\cite{Togashi2001,Ohkubo2008,Biancalani2012}, or the dynamic Ising
model \cite{Glauber1963} for a set of fully connected spins. The
general properties of this stochastic system, and its application
to population genetics in fluctuating environment were discussed by
Horsthemke and Lefeve \cite{Horsthemke}. 

The behavior of this system is markedly different from its mean field,
deterministic approximation. Indeed, the equation for $\left\langle n\right\rangle $,
the mean number of individuals in one state, is:
\begin{equation}
\frac{d\left\langle n\right\rangle }{dt}=\left\langle W^{+}(n)-W^{-}(n)\right\rangle =\epsilon\left(N-2\left\langle n\right\rangle \right)\label{eq:mean}
\end{equation}
and has a stable stationary solution $\left\langle n\right\rangle =N/2$.
However, for small values of $\epsilon$, \emph{i.e.} $\epsilon\ll1/N$,
the system is observed most of time in one the two boundary states
$n=0$ or $n=N$, and seldom in states close to $n=N/2$. The bistability
of the system is caused solely by the noise and cannot be captured
by the mean field equation (\ref{eq:mean}). 

The reason behind the bistability is the following: in the absence
of spontaneous switching ($\epsilon=0$), the states $n=0$ (all individuals
in configuration $B$) and $n=N$ (all individuals in configuration
$A$) are absorbing: $W^{+}(0)=W^{-}(N)=0$. Eventually, the system
will end up in one of these two states and remain there. When $\epsilon>0$,
these states cease to be absorbing. However, the mean residence time
$\tau$ in these states is $\left(W^{+}(\alpha)+W^{-}(\alpha)\right)^{-1}=1/\epsilon N$
(where $\alpha=0,N$) while the residence time in other states is
$O(1)$. Therefore, in the regime $\epsilon N\ll1$, the system is
observed mostly in the boundary states. 

In their article, Biancalani et al.  computed $T(0)$, the mean switching
time (the mean first passage time) from state $n=0$ to state $n=N$,
and show that the observation of this quantity can lead to the measure
of the parameter $\epsilon$ of this stochastic system. For this
computation, they expanded the Master equation of the stochastic system
in powers of $1/N$ and neglected terms of $O(1/N^{3})$ to obtain
the forward and backward Fokker-Plank equation, from which the mean
switching time can be obtained ( \cite{Biancalani2014} , equation
(4) and Supplementary Materials, equations (4) and (11) ). This approximation
is fragile, specially for small $N$ where the noise is strong. In
particular, to compute $T(0)$, they have used two different approximations,
one of which is valid for $0.2\lesssim N\epsilon$ and the other for
$N\epsilon\rightarrow0$, and there is no clear criterion for their
overlap. In this article, we compute the exact expression for $T(0)$
without any approximation, which is valid for all values of $\epsilon$.
We further extend this investigation by giving the exact solution
of the discrete Master equation through the use of the probability
generating function associated to the probabilities. Other quantities
that we compute, such as the dynamics of the moments or the dynamics
of the boundary states probabilities, provide other useful tools to
measure and investigate this system.

This article is organized as follow: in the next section, we give
the exact expression for the mean first passage time $T(n)$. The
following section is devoted to the solution of the Master equation.
The final section is devoted to discussion and conclusion.

\section{Switching time.}

Preparing the system at time $t=0$ in the initial state $n=m$, the
system evolves and will reach the state $n=N$ for the first time
at some time $T(m)$. The mean first passage times $\bar{T}(m)$ are
obtained from the backward Kolmogorov equation and form the linear
system \cite{Gardiner2004}
\begin{eqnarray}
W^{+}(0)\left(\bar{T}(1)-\bar{T}(0)\right) & = & -1\label{eq:T0}\\
W^{+}(m)\left(\bar{T}(m+1)-\bar{T}(m)\right) & +\nonumber \\
W^{-}(m)\left(\bar{T}(m-1)-\bar{T}(m)\right) & = & -1\label{eq:Tm}
\end{eqnarray}
where $0<m<N$. Note that as $W^{-}(0)=0$, we don't need to write
a separate equation (\ref{eq:T0}) for the boundary term $\bar{T}(0)$
; the above notation however is clearer and highlights the boundary
condition. Note also that by definition, $\bar{T}(N)=0$, so the above
square system of linear equations is well posed. 

\begin{figure}
\includegraphics[width=1\columnwidth]{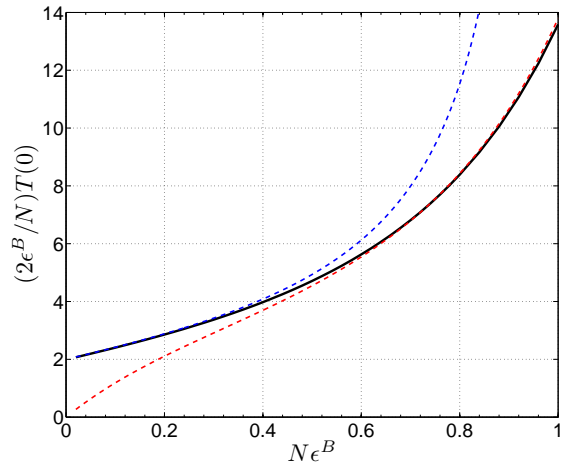}

\caption{(Color online) Exact result for the first first passage time (solid
line, black) as a function of $N$ for $\epsilon^{\mbox{B}}=1/500$,
$r=1$ and its comparison to the two solutions provided by Biancalani
et al.(\cite{Biancalani2014}, Figure 5) : dotted curve, blue for
$\epsilon\rightarrow0$ ; dotted curve, red for $N\epsilon^{\mbox{B}}\gtrsim0.5$.
\label{fig:comparison}}

\end{figure}

Using the continuous approximation $n\rightarrow x=n/N$, $\bar{T}(m)\rightarrow\bar{t}(x)$,
and developing equation (\ref{eq:Tm}) to the second order in $(1/N)$,
one obtains the second order differential equation for $\bar{t}(x)$
which can be solved in terms of the hypergeometric function, as was
done by Biancalani et al\cite{Biancalani2014} (see \ref{sub:biancalani}).
The continuous limit is however fragile when $\epsilon\rightarrow0$,
and the first solution obtained by Biancalani et al. does not converge
to the right value in this limit. This is due to the absorbing boundary
condition $t'(0)=0$ used in the continuous approximation, which fails
in the limit $\epsilon\rightarrow0$ as it can be observed directly
from equation (\ref{eq:T0}) (see also \cite{Biancalani2014} Supplement.
Materials). In order to resolve this problem, they have resorted to
a limit process for the case $\epsilon\rightarrow0$ by approximating
(\cite{Biancalani2014} Supplement. Materials, eq.(28) ) 
\[
_{2}F_{1}(\frac{1}{2},u;\frac{3}{2};\frac{1}{1+2\epsilon})\approx{}_{2}F_{1}(\frac{1}{2},u;\frac{3}{2};1)
\]
where $u=N\epsilon$ or $1-N\epsilon$\emph{, i.e.} setting $\epsilon=0$
in the fourth argument of the hypergeometric function, but not in
the second. This \emph{ad hoc} approximation gives the correct solution
for $\epsilon\rightarrow0$; no criterion however can be obtained
for the overlap between the two solutions (figure \ref{fig:comparison}). 

These complications are due to the continuous approximation and can
be avoided if the solution is computed directly for the discrete equations
(\ref{eq:T0},\ref{eq:Tm}). The discrete solution is computationally
much simpler, is valid for the whole range of $\epsilon$ and $N$
and does not involve any approximation; specifically, the boundary
conditions are set naturally and don't need to be adjusted as a function
of $\epsilon$. The solution is obtained by setting $y_{k}=\bar{T}(k)-\bar{T}(k-1)$,
which transforms equations (\ref{eq:T0},\ref{eq:Tm}) into a simple
\emph{one-term} recurrence equation. The exact solution is then 
\[
y_{k+1}=-\sum_{i=0}^{k}\frac{(N-k+\epsilon)_{(k-i)}}{(N-k)_{(k-i+1)}}\frac{(i+1)_{(k-i)}}{(i+\epsilon)_{(k-i+1)}}\,\,\,0\le k<N
\]
where $(\alpha)_{(m)}=\alpha(\alpha+1)...(\alpha+m-1)=\Gamma(\alpha+m)/\Gamma(\alpha)$
is the Pochhammer symbol. 

\begin{figure}
\begin{centering}
\includegraphics[width=0.9\columnwidth]{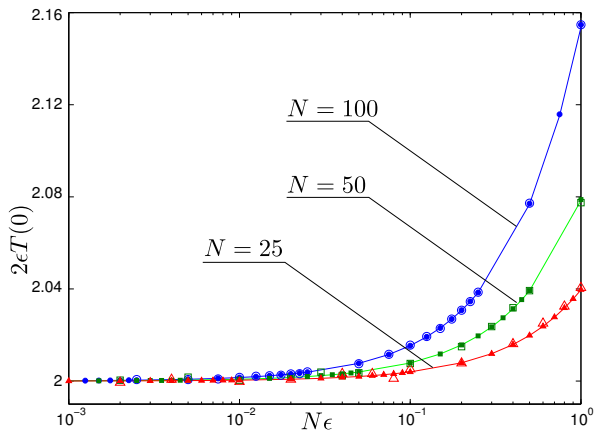}
\par\end{centering}

\caption{(Color online) Switching time as a function of $\epsilon$ for three
different values of $N$. Empty symbols: Numerical simulation by a
Gillespie algorithm over $10^{7}$paths ; filled symbols: numerical
solution of the linear system (\ref{eq:Tm}-\ref{eq:T0}); Solid lines:
theoretical expression (\ref{eq:first order}). \label{fig:T0}}
\end{figure}

As $\bar{T}(N)=0$, the first passage times $\bar{T}(m)$ are easily
recovered from the $y_{k}$ :
\[
\bar{T}(m)=-\sum_{k=m}^{N-1}y_{k+1}
\]
In particular, the mean time to move from one boundary state to the
other is
\begin{equation}
\bar{T}(0)=\sum_{k=0}^{N-1}\sum_{i=0}^{k}\frac{(N-k+\epsilon)_{(k-i)}}{(N-k)_{(k-i+1)}}\frac{(i+1)_{(k-i)}}{(i+\epsilon)_{k-i+1}}\label{eq:T0solved}
\end{equation}
The above expression is computationally simpler than the product of
two hypergeometric functions and involves only simple, finite arithmetics.
Its expansion in the first two powers of $\epsilon$ gives (see Mathematical
Details):
\begin{equation}
\bar{T}(0)=\frac{1}{\epsilon}+2\frac{N-1}{N}+O(\epsilon)\label{eq:first order}
\end{equation}
Figure \ref{fig:T0} shows the remarkable accuracy of this formula
for $N\epsilon\in[0,1]$ and $N\lesssim100$, \emph{i.e. }the relevant
range where bi-stability can be observed.
 The analysis can be extended to compute the linear term in $\epsilon$
in equation (\ref{eq:first order}) (see section \ref{sec:Mathematical-details}.A)

Equations (\ref{eq:T0solved},\ref{eq:first order}) have been obtained
by setting $r=1$, \emph{i.e. }by scaling time and $\epsilon$ by
the factor $r$. Restoring the non-scaled time ($t\rightarrow t/r$,
$\epsilon\rightarrow\epsilon/r$), we have 
\[
\bar{T}_{\epsilon,r}^{(\mbox{ns)}}(0)=\frac{1}{r}\bar{T}_{\epsilon/r}(0)
\]
and in particular, the leading terms of the development are
\[
\bar{T}_{\epsilon,r}^{(\mbox{ns)}}(0)=\frac{1}{\epsilon}+\frac{2}{r}\frac{N-1}{N}+\frac{1}{r}O(\frac{\epsilon}{r})
\]
Therefore, it is possible in principle, by measuring the switching
time for different system size $N$, to measure independently the
parameters $\epsilon$ and $r$. 

Note that the rate coefficients used by Biancalani et al. are given
in terms of proportions, \emph{i.e. $r^{\mbox{B}}=N^{2}r$ }and $\epsilon^{\mbox{B}}=N\epsilon$.
Figure \ref{fig:comparison} shows the comparison between our exact
result and the Biancalani et al. approximate solutions when this scaling
is taken into account, for the full range of $N\epsilon$. It can
be observed that the two solutions obtained by Biancalani et al. and
their overlap can be recovered from the exact solution we provide
here. 

In a yet unpublished article, Saito and Kaneko \cite{Saito2014} have
also computed the switching time for this stochastic system. Their
method consists in obtaining an approximation for the \emph{residence}
time $t_{0,j}$ in each state $j$ beginning from state 0 and then
summing up these residence times to obtain the switching time. Their
analytical result for the switching time has a very different \emph{form}
that the relation (\ref{eq:T0solved}) and doesn't seem amenable to
easy computation of the interesting limiting case $N\epsilon\ll1$.
However, their formula produces the same numerical results than the
relation (\ref{eq:T0solved}) of this article.

\section{Solving the master equation.}

\begin{figure}
\begin{centering}
\includegraphics[width=1\columnwidth]{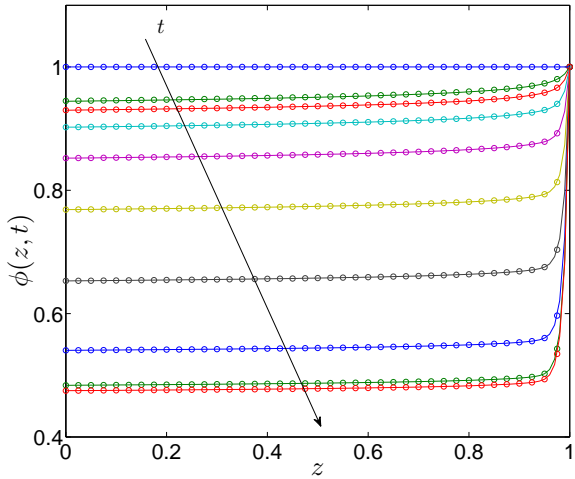}
\par\end{centering}

\caption{(Color online) The PGF function $\phi(z,t)$ as a function of $z$
at times $t\in\{0,1,2,4,8,16,32,64,128,256\}/(128\epsilon)$ for $N=100$
and $\epsilon=0.01$. Solid lines: theoretical expression (\ref{eq:PGFsolved}).
Circles: solution obtained by the numerical resolution of the Master
equation (\ref{eq:master}) and computation of its PGF.\label{fig:pfg}}

\end{figure}

The mean first passage is one tool to study the stochastic system
described by the transition rates (\ref{eq:Wm},\ref{eq:Wp}). A \emph{complete}
description can be obtained by solving directly the master equation
governing the probabilities $P(n,t)$ to observe $n$ individuals
in state $A$ at time $t$:
\begin{eqnarray}
\frac{\partial P(n,t)}{\partial t} & = & W^{+}(n-1)P(n-1,t)-W^{+}(n)P(n,t)\nonumber \\
 & + & W^{-}(n+1)P(n+1,t)-W^{-}(n)P(n,t)\label{eq:master}
\end{eqnarray}
We note that the above stochastic system does \emph{not need} a moment
closure approximation, \emph{i.e.} the equation for the $k$th moment
involves only moments of order lower than $k$. Therefore, a hierarchical
system of equations can be established to derive all the moments of
this system. The probability generating function is a powerful tool
to investigate such Master equations \cite{Gardiner2004,VanKampen1992}.
The PGF is defined as 

\[
\phi(z,t)=\left\langle z^{n}\right\rangle =\sum_{n=0}^{N}P(n,t)z^{n}
\]
and contains the most complete information we can have on the given
stochastic process: all the moments and probabilities can be obtained
from its derivatives at either $z=1$ or $z=0$. The equation governing
the PGF can be extracted from the master equation (\ref{eq:master})
(see section \ref{sub:PGF}) and reads:
\begin{eqnarray}
\frac{\partial\phi}{\partial t} & = & -z(z-1)^{2}\frac{\partial^{2}\phi}{\partial z^{2}}\nonumber \\
 & + & (z-1)\left[\left(N-1-\epsilon\right)z-\left(N-1+\epsilon\right)\right]\frac{\partial\phi}{\partial z}\nonumber \\
 & + & \epsilon N(z-1)\phi\label{eq:PGF}
\end{eqnarray}
The solution of equation (\ref{eq:PGF}) can be exactly computed (see
section \ref{sub:PGF}) as the superposition of polynomial eigenfunctions
\begin{equation}
\phi(z,t)=\sum_{n=0}^{N}C_{n}\phi_{n}(z)e^{\lambda_{n}t}\label{eq:PGFsolved}
\end{equation}
where the eigenvalues are 
\[
\lambda_{n}=-n(n-1+2\epsilon),
\]
the eigenfunctions are polynomials in $z$ 
\[
\phi_{n}(z)=\sum_{k=n}^{N}a_{k}^{n}(1-z)^{k}
\]
and the coefficients $C_{n}$ depend on the initial condition. The
initial condition we use here is the same as in the previous section,
$i.e.$ $P(n,0)=\delta_{n,0}$ which implies that $\phi(z,0)=1$.
The exact expression for the coefficients $a_{k}^{n}$ , $C_{n}$
and their product are given in the section \ref{sub:PGF}. The agreement
between the solution (\ref{eq:PGFsolved}) and the direct numerical
solution of the Master equation is displayed in figure \ref{fig:pfg}.

The PGF contains the most complete information on the stochastic process
under investigation. Some quantities of interest extracted from it
are given below.

\subsection{Stationary probabilities.}

The stationary probabilities attained at large times are 
\begin{figure}
\begin{centering}
\includegraphics[width=1\columnwidth]{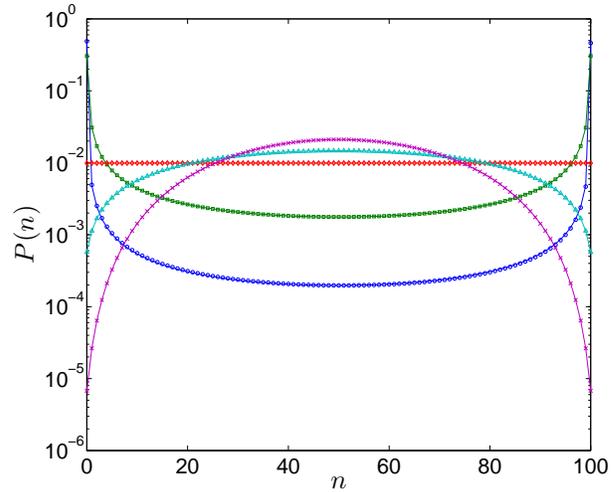}
\par\end{centering}

\caption{(Color online) The stationary probabilities $P(n,\infty)$ as a function
of $n$ for $N=100$ and various $\epsilon$. Solid lines: exact expression
(\ref{eq:probastat}), symbols: numerical resolution of the Master
equation. $\epsilon=0.01$ (blue circles), $0.1$ (green squares),
1 (red diamonds), 2 (diamonds, cyan) and 4 ($\times$, purple). \label{fig:stationaryprobabilities}}

\end{figure}
 
\begin{equation}
P(n,\infty)=\binom{N}{n}\frac{(\epsilon)_{(n)}(\epsilon)_{(N-n)}}{(2\epsilon)_{(N)}}\label{eq:probastat}
\end{equation}
(see section \ref{sub:PGF}) and their comparison to numerical solution
of the Master equation is displayed in figure \ref{fig:stationaryprobabilities}.
Note the qualitative change of behavior at $\epsilon=1$. Expression
(\ref{eq:probastat}) is equivalent to the expression found by Biancalani
et \emph{al.} \cite{Biancalani2014} in the continuous approximation,
with the advantage of being well defined for all $n$, including $n=0,N$.
In particular, for $\epsilon N\ll1$,
\[
P(n,\infty)\begin{cases}
(1-H_{N-1}\epsilon)/2+O(\epsilon^{2}) & n=0,N\\
\frac{N\epsilon}{2n(N-n)}+O(\epsilon^{2}) & n\ne0,N
\end{cases}
\]
where $H_{m}$ is the harmonic number $\sum_{i=1}^{m}i^{-1}$.

\subsection{Factorial moments.}

For the purposes of experimental measurements of the parameters, other
dynamical quantities can be of interest. The most robust of these
quantities are the factorial moments 
\[
\left\langle (n,q)\right\rangle =\left\langle n(n-1)...(n-q+1)\right\rangle 
\]
where $(n,q)$ is used to denote the decreasing Pochhammer symbol.
The factorial moments are obtained by successive derivation of the
PGF
\begin{eqnarray}
\left\langle (n,q)\right\rangle  & =q! & \left.\frac{\partial^{q}\phi}{\partial z^{q}}\right|_{z=1}\nonumber \\
 & = & (-1)^{q}q!\sum_{i=0}^{q}C_{i}a_{q}^{i}e^{\lambda_{i}t}\label{eq:factorialmoment}
\end{eqnarray}
Note that the $q$th factorial moment involves only $q+1$ eigenfunctions.
The two first factorial moments are 
\begin{eqnarray*}
\left\langle n\right\rangle  & = & \frac{N}{2}\left(1-e^{-2\epsilon t}\right)\\
\left\langle n(n-1)\right\rangle  & = & \frac{N(N-1)}{2}\\
 & \times & \left(\frac{1+\epsilon}{1+2\epsilon}-e^{-2\epsilon t}+\frac{\epsilon}{1+2\epsilon}e^{-2(1+2\epsilon)t}\right)
\end{eqnarray*}
For $N\epsilon\ll1$, only the two first terms in the sum (\ref{eq:factorialmoment})
contribute significantly to the factorial moments for $t\gtrsim1$.
In particular, for large times, 
\[
\left\langle (n,q)\right\rangle \rightarrow(N,q)\frac{1-H_{q-1}\epsilon}{2}
\]

\subsection{Equilibrium time.}

\begin{figure}
\begin{centering}
\includegraphics[width=1\columnwidth]{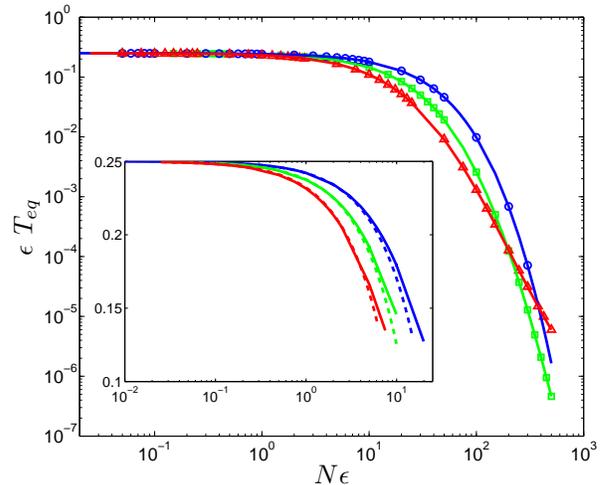}
\par\end{centering}

\caption{(Color online) $T_{eq}$ as a function of $\epsilon$ for different
values of $N$. Solid lines: theoretical expression (\ref{eq:teq2})
; symbols: numerical resolution of the master equation (blue circles
$N=100$; green squares $N=50$ ; red triangles $N=25$). Inset: comparison
between the exact expression (\ref{eq:teq2}) (solid lines) and its
approximation (\ref{eq:teqapprox})(dashed lines) for $N\epsilon\lesssim1$
and $N=100,50$ and $25$. \label{fig:teq}}

\end{figure}

Finally, we can define an equilibrium time $T_{eq}$ by studying the
dynamics of the decrease in $P(0,t)$ or increase in $P(N,t)$. The
measure we choose to use here is 
\begin{equation}
T_{eq}=\int_{0}^{\infty}\left\{ P(N,\infty)-P(N,t)\right\} dt\label{eq:Teq}
\end{equation}
which is a generalization of the mean first passage time (see \ref{sub:PGF}
). The expressions for the two boundary probabilities are found to
be 
\begin{eqnarray*}
P(0,t) & = & \sum_{n=0}^{N}(-)^{N-n}C_{n}a_{N}^{n}e^{\lambda_{n}t}\\
P(N,t) & = & (-)^{N}\sum_{n=0}^{N}C_{n}a_{N}^{n}e^{\lambda_{n}t}
\end{eqnarray*}
and therefore
\begin{eqnarray}
T_{eq} & = & (-)^{N}\sum_{n=1}^{N}C_{n}a_{N}^{n}/\lambda_{n}\label{eq:teq2}
\end{eqnarray}
For $N\epsilon\lesssim1$, eq.(\ref{eq:teq2}) is approximated by
\begin{equation}
T_{eq}=\frac{1}{4\epsilon}-\frac{1}{4}\left(H_{N-1}-2+\frac{2}{N}\right)\label{eq:teqapprox}
\end{equation}
Figure \ref{fig:teq} displays $T_{eq}$ as a function of $\epsilon$
and its comparison to numerical solution of the master equation.

\section{Conclusion.}

As discussed in the introduction, noise induced bi-stability has been
intensely investigated, specially in genetic networks. In general,
the chemical Master equations are too complex to be solved exactly
and various approximation techniques have been developed to tackle
this problem. In some cases, exact analytical solutions have been
obtained using the probability generating function. Shahrezaei and
Swain \cite{Shahrezaei2008} have studied a three stage model of simple
gene expression (DNA state, RNA, Protein) and obtained the protein
number distribution. Grima et al. \cite{Grima2012} have investigated
the steady state distribution of a two component (DNA state, Protein)
genetic feedback loop and have been able to obtain exact analytical
results using the PGF technique. In the first case, the PGF equation
is a first order partial differential equation and can be solved by
the method of characteristics. In the second case, the model can be
reduced to two coupled one component systems and the PGF equation
reduced to two ordinary coupled first order differential equations.
Chemical Master equations analogous to these cases could in principle
be investigated with the same technique. 

In this work, we have extended the investigation by Biancalani et
\emph{al. \cite{Biancalani2014} }of another noise induced bistable
system which belongs to the second class of models discussed above.
First, we have obtained the exact solution for the mean first passage
time which is the main result of the above cited article. Second,
we have solved the full master equation associated with this system
and obtained other useful quantities for parameter estimations of
such systems. We have obtained these results for the original, discrete
system without resorting to the Taylor expansion of the Master equation
in powers of $1/N$. Discrete solutions have the advantage of being
clearly defined and avoid spurious effect happening at the boundaries,
specially for the interesting case of small $\epsilon$. Moreover,
these solutions involve only simple arithmetic and are easily computed.

\section{Mathematical details.\label{sec:Mathematical-details}}

\subsection{Series expansion of the exact solution of the switching time.}

The exact solution (\ref{eq:T0solved}) contains a double sum, where
only the terms $i=0$ contain $\epsilon^{-1}$ factors. Separating
these two contributions, the solution becomes:
\begin{eqnarray*}
\bar{T}(0) & = & \frac{1}{N\epsilon}\sum_{k=0}^{N-1}\frac{(1)_{k}}{(1+\epsilon)_{k}}\frac{(N-k+\epsilon)_{k}}{(N-k)_{k}}\\
 & + & \sum_{k=1}^{N-1}\sum_{i=1}^{k}\frac{(N-k+\epsilon)_{(k-i)}}{(N-k)_{(k-i+1)}}\frac{(i+1)_{(k-i)}}{(i+\epsilon)_{k-i+1}}
\end{eqnarray*}
Expanding the first sum to the first order in $\epsilon$ necessitates
only simple expansion in factors of the form $m/(m+\epsilon)=1-\epsilon/m+O(\epsilon^{2})$
and leads to 
\[
\frac{1}{\epsilon}-H_{N-1}+2\frac{N-1}{N}
\]
where the Harmonic number $H_{m}=\sum_{i=1}^{m}(1/i)$. Evaluating
the second sum for $\epsilon=0$ results in 
\[
\sum_{k=1}^{N-1}\sum_{i=1}^{k}\frac{1}{i(N-i)}=H_{N-1}
\]
Adding the two contributions results in (eq.\ref{eq:first order}):
\[
\bar{T}(0)=\frac{1}{\epsilon}+2\frac{N-1}{N}
\]
The next term in the series expansion of $\bar{T}(0)$ is found to
be 
\[
-\frac{2\epsilon}{N}\left(H_{N-1}+NH_{N-1}^{(2)}-2(N-1)\right)
\]
Note that algorithmically, the computation of $\bar{T}(0)$ (expression
(\ref{eq:T0solved}) ) necessitates only the calculation of $N$ ratios
of the form $(m+1)/(m+\epsilon)$ and $(m+\epsilon)/m$ which can
be stored in an array. The $\bar{T}(0$) involves then only multiplications
and sums of these elements. The Hypergeometric function on the other
hand is defined as 
\[
_{2}F_{1}(a,b;c;z)=\sum_{n=0}^{\infty}\frac{(a)_{(n)}(b)_{(n)}}{(c)_{(n)}}\frac{z^{n}}{n!}
\]
and its efficient implementation requires specific algorithms.

\subsection{Solution of Biancalani et al. for the switching time.\label{sub:biancalani}}

In non scaled time, the Biancalani et al. solution is 
\begin{eqnarray*}
\bar{T}^{\mbox{ns}}(0) & = & \frac{1}{r'}\frac{2N^{2}}{1+2\epsilon'/r'}\,_{2}F_{1}\left(\frac{1}{2},1-N\frac{\epsilon'}{r'};\frac{3}{2};\frac{1}{1+2\epsilon'/r'}\right)\\
 & \times & \,_{2}F_{1}\left(\frac{1}{2},N\frac{\epsilon'}{r'};\frac{3}{2};\frac{1}{1+2\epsilon'/r'}\right)
\end{eqnarray*}
where the rates $\epsilon'$ and $r'$ are related to the rates $\epsilon$,$r$
used in this article through:
\[
\epsilon'=N\epsilon\,\,;\,\, r'=N^{2}r
\]

\subsection{Deriving and solving the PGF equation.\label{sub:PGF}}

\paragraph{PGF.}

The equation for the evolution of the PGF is obtained by multiplying
the master equation(\ref{eq:master}) by $z^{n}$ and summing over
$n$ \cite{Houchmandzadeh2010}. This operation leads to 
\begin{equation}
\frac{\partial\phi}{\partial t}=\left\langle (z^{n+1}-z^{n})W^{+}(n)\right\rangle +\left\langle (z^{n-1}-z^{n})W^{-}(n)\right\rangle \label{eq:pfgderivation}
\end{equation}
The rates $W^{\pm}(n)$ are polynomials of second degree in $n$ and
by the definition of the PGF, 
\[
\left\langle n^{r}z^{n}\right\rangle =\left(z\frac{\partial}{\partial z}\right)^{r}\phi
\]
Application of the above rule to equation (\ref{eq:pfgderivation})
leads to equation (\ref{eq:PGF}).

\paragraph{Eigenfunctions.}

Equation (\ref{eq:PGF}) can be transformed into a hypergeometric
equation by a change of variable $x=(z-1)^{-1}$. It is however much
simpler to use the fact that by definition, the function $\phi(z,t)$
is a polynomial of degree $N$ in $z$ and search for the eigenfunctions
of equation (\ref{eq:PGF}) in term of polynomials of the following
form: 
\[
\phi_{n}(z)=\sum_{k=0}^{N}a_{k}^{n}(1-z)^{k}
\]
\emph{i.e. 
\[
\phi(z,t)=\sum_{n=0}^{N}C_{n}\phi_{n}(z)e^{\lambda_{n}t}
\]
}Insertion of these polynomials into equation (\ref{eq:PGF}) shows
that non-trivial solutions (\emph{i.e.} $\ne0$) are possible only
for the eigenvalues 
\[
\lambda_{n}=-n(n-1+2\epsilon)\,\,\,\, n=0,1,...,N
\]
which leads to a one term recurrence relation on the coefficients
$a_{k}^{n}$ : 
\begin{eqnarray*}
a_{k}^{n} & = & 0\,\,\,\,\,\,(k<n)\\
a_{n}^{n} & = & 1\\
a_{k+1}^{n} & = & -\frac{(N-k)(k+\epsilon)}{(k+1)(k+2\epsilon)-n(n-1+2\epsilon)}a_{k}^{n}\,\,\,\,\,\,(n\le k<N)
\end{eqnarray*}
As it can be noticed, $\phi_{n}$ is written as polynomial in powers
of $(1-z)$ and not $z$. This choice is not arbitrary: it is this
change of variable which allows to obtain a one term recurrence relation
between the coefficients $a_{k}^{n}$. Writing $\phi_{n}$ as a polynomial
in $z$ leads to a two terms recurrence relation which is much more
intricate to solve exactly. 

The coefficients $a_{k}^{n}$ can be computed in explicit forms:
\begin{equation}
a_{k}^{n}=(-)^{k-n}\binom{N-n}{k-n}\frac{(\epsilon+n)_{(k-n)}}{(2\epsilon+2n)_{(k-n)}}\,\,\,\,\,\,(n\le k<N)\label{eq:ank}
\end{equation}
Alternatively, the eigenfunctions can also be given in terms of the
hypergeometric function:
\begin{equation}
\phi_{n}(z)=(1-z)^{n}\,\,_{2}F_{1}(n-N,n+\epsilon;2n+2\epsilon;1-z)\label{eq:2F1}
\end{equation}
The amplitudes $C_{n}$ depend on the initial condition. For $P(n,0)=\delta_{n,0}$
and therefore $\phi(z,0)=1$, the amplitudes obey the triangular linear
system
\begin{eqnarray*}
C_{0} & = & 1\\
\sum_{n=0}^{k}C_{n}a_{k}^{n} & = & 0\,\,\,\,(k>0)
\end{eqnarray*}
which can be explicitly solved 
\begin{equation}
C_{n}=\left(\begin{array}{c}
N\\
n
\end{array}\right)\frac{(\epsilon)_{(n)}}{(2\epsilon+n-1)_{(n)}}\label{eq:Cn}
\end{equation}
and therefore, 
\[
C_{n}a_{k}^{n}=(-)^{k-n}\binom{N}{k}\binom{k}{n}\frac{(\epsilon)_{(k)}}{(2\epsilon+n)_{(k)}}\frac{2\epsilon+2n-1}{2\epsilon+n-1}
\]

\paragraph{Stationary probabilities.}

As all eigenvalues except $\lambda_{0}$ are negative, for large times
the PGF is simply 
\[
\phi(z)={}_{2}F_{1}(-N,\epsilon;2\epsilon;1-z)
\]
where we have used the hypergeometric representation (eq. \ref{eq:2F1})
of the eigenfunctions. Using the relations 
\begin{eqnarray*}
_{2}F_{1}(-m,b;c;1) & = & \frac{(c-b)_{(m)}}{(c)_{(m)}}\\
\frac{d^{n}}{dz^{n}}\,_{2}F_{1}(a,b;c;z) & = & \frac{(a)_{(n)}(b)_{(n)}}{(c)_{(n)}}\,_{2}F_{1}(a+n,b+n;c+n;z)
\end{eqnarray*}
we obtain 
\begin{eqnarray}
P(n) & = & \frac{1}{n!}\left.\frac{d^{n}\phi}{dz^{n}}\right|_{z=0}\nonumber \\
 & = & (-1)^{n}\frac{(-N)_{(n)}}{n!}\frac{(\epsilon)_{(n)}}{(2\epsilon)_{(n)}}\frac{(\epsilon)_{(N-n)}}{(2\epsilon+n)_{(N-n)}}\label{eq:stationary}
\end{eqnarray}
As 
\[
(2\epsilon)_{(n)}(2\epsilon+n)_{(N-n)}=(2\epsilon)_{N}
\]
we recover the relation (\ref{eq:probastat}) on the stationary probabilities.

\paragraph{Factorial moments.}

Using the above expression, the factorial moments are 
\[
\left\langle (n,q)\right\rangle =(N,q)\sum_{i=0}^{q}(-)^{i}\binom{q}{i}\frac{(\epsilon)_{(q)}}{(2\epsilon+i)_{(q)}}\frac{2\epsilon+2i-1}{2\epsilon+i-1}e^{\lambda_{i}t}
\]

\paragraph{Equilibrium times.}

Many different measures can be used for the equilibrium time of the
system. The expression we use
\begin{equation}
T_{eq}=\int_{0}^{\infty}\left(P(N,\infty)-P(N,t)\right)dt\label{eq:teqapp}
\end{equation}
is the extension of the mean time to absorption to the case when the
boundary state is not absorbing. The reason is the following: If the
state $N$ were the only absorbing state, whatever the initial condition
$m$, $P(N,t)\rightarrow1$ as $t\rightarrow\infty$. The probability
of survival until time $T$, beginning in the state $m$ is 
\[
Q(m,T)=1-P(N,T)
\]
and the probability density of not being absorbed during $[T,T+dt]$
is therefore $-\partial_{T}Q(m,T)$. Therefore, the mean time to absorption
is 
\begin{eqnarray*}
\bar{T}(m) & = & -\int_{0}^{\infty}T\partial_{T}Q(m,T)dT\\
 & = & \int_{0}^{\infty}\left(1-P(N,T)\right)dT\\
 & = & \int_{0}^{\infty}\left(P(N,\infty)-P(N,T)\right)dT
\end{eqnarray*}
We see that in the case of an absorbing state $N$, our definition
of $T_{eq}$ and the mean time to absorption are the same. We continue
to use $T_{eq}$ as a measure of the equilibrium time when $N$ is
not absorbing.

\paragraph{Probabilities. }

The probabilities are extracted from the PGF by collecting the coefficients
of powers of $z$:
\[
P(n,t)=\sum_{k=0}^{N}b_{k}^{n}\exp(\lambda_{k}t)
\]
where 
\[
b_{k}^{n}=(-)^{n}C_{k}\sum_{j=k}^{N}\left(\begin{array}{c}
j\\
n
\end{array}\right)a_{j}^{k}.
\]

\bibliographystyle{unsrt}

\begin{thebibliography}{}

\end{thebibliography}


\begin{thebibliography}{10}

\bibitem{Vilar2002}
Jos\'{e} M~G Vilar, Hao~Yuan Kueh, Naama Barkai, and Stanislas Leibler.
\newblock {Mechanisms of noise-resistance in genetic oscillators.}
\newblock {\em Proceedings of the National Academy of Sciences of the United
  States of America}, 99(9):5988--92, April 2002.

\bibitem{Houchmandzadeh2012a}
Bahram Houchmandzadeh and Marcel Vallade.
\newblock {Selection for altruism through random drift in variable size
  populations.}
\newblock {\em BMC Evol Biol}, 12:61, 2012.

\bibitem{Korolev2010}
K~S Korolev, Mikkel Avlund, Oskar Hallatschek, and David~R Nelson.
\newblock {Genetic demixing and evolution in linear stepping stone models.}
\newblock {\em Reviews of modern physics}, 82(2):1691--1718, June 2010.

\bibitem{Houchmandzadeh2008}
B~Houchmandzadeh.
\newblock {Neutral clustering in a simple experimental ecological community.}
\newblock {\em Phys Rev Lett}, 101(7):78103, 2008.

\bibitem{Dumonteil2014}
Eric Dumonteil, Fausto Malvagi, Andrea Zoia, Alain Mazzolo, Davide Artusio,
  Cyril Dieudonn\'{e}, and Cl\'{e}lia {De Mulatier}.
\newblock {Particle clustering in Monte Carlo criticality simulations}.
\newblock {\em Annals of Nuclear Energy}, 63:612--618, January 2014.

\bibitem{Horsthemke}
W.~Horsthemke and R.~Lefeve.
\newblock {\em {Noise-Induced Transitions: Theory and Applications in Physics,
  Chemistry, and Biology}}.
\newblock Springer-Verlag, Berlin, 1986.

\bibitem{Samoilov2005}
Michael Samoilov, Sergey Plyasunov, and Adam~P Arkin.
\newblock {Stochastic amplification and signaling in enzymatic futile cycles
  through noise-induced bistability with oscillations.}
\newblock {\em Proceedings of the National Academy of Sciences of the United
  States of America}, 102(7):2310--5, February 2005.

\bibitem{Artyomov2007}
Maxim~N Artyomov, Jayajit Das, Mehran Kardar, and Arup~K Chakraborty.
\newblock {Purely stochastic binary decisions in cell signaling models without
  underlying deterministic bistabilities.}
\newblock {\em Proceedings of the National Academy of Sciences of the United
  States of America}, 104(48):18958--63, November 2007.

\bibitem{Qian2009}
Hong Qian, Pei-Zhe Shi, and Jianhua Xing.
\newblock {Stochastic bifurcation, slow fluctuations, and bistability as an
  origin of biochemical complexity.}
\newblock {\em Physical chemistry chemical physics : PCCP}, 11(24):4861--70,
  June 2009.

\bibitem{Thomas2014}
Philipp Thomas, Nikola Popovi\'{c}, and Ramon Grima.
\newblock {Phenotypic switching in gene regulatory networks.}
\newblock {\em Proceedings of the National Academy of Sciences of the United
  States of America}, 111(19):6994--9, May 2014.

\bibitem{To2010}
Tsz-Leung To and Narendra Maheshri.
\newblock {Noise can induce bimodality in positive transcriptional feedback
  loops without bistability.}
\newblock {\em Science (New York, N.Y.)}, 327(5969):1142--5, February 2010.

\bibitem{Biancalani2014}
Tommaso Biancalani, Louise Dyson, and Alan~J. McKane.
\newblock {Noise-Induced Bistable States and Their Mean Switching Time in
  Foraging Colonies}.
\newblock {\em Physical Review Letters}, 112(3):038101, January 2014.

\bibitem{Moran1962}
P~A~P Moran.
\newblock {\em {The Statistical processes of of evolutionary theory}}.
\newblock Oxford University Press, 1962.

\bibitem{Togashi2001}
Yuichi Togashi and Kunihiko Kaneko.
\newblock {Transitions Induced by the Discreteness of Molecules in a Small
  Autocatalytic System}.
\newblock {\em Physical Review Letters}, 86(11):2459--2462, March 2001.

\bibitem{Ohkubo2008}
Jun Ohkubo, Nadav Shnerb, and David {A. Kessler}.
\newblock {Transition Phenomena Induced by Internal Noise and Quasi-Absorbing
  State}.
\newblock {\em Journal of the Physical Society of Japan}, 77(4):044002, April
  2008.

\bibitem{Biancalani2012}
Tommaso Biancalani, Tim Rogers, and Alan~J. McKane.
\newblock {Noise-induced metastability in biochemical networks}.
\newblock {\em Physical Review E}, 86(1):010106, July 2012.

\bibitem{Glauber1963}
Roy~J. Glauber.
\newblock {Time-Dependent Statistics of the Ising Model}.
\newblock {\em Journal of Mathematical Physics}, 4(2):294, December 1963.

\bibitem{Gardiner2004}
C~Gardiner.
\newblock {\em {Handbook of Stochastic Methods: for Physics, Chemistry and the
  Natural Sciences}}.
\newblock Springer, 2004.

\bibitem{Saito2014}
Nen Saito and Kunihiko Kaneko.
\newblock {Theoretical Analysis of Discreteness-Induced Transition in
  Autocatalytic Reaction Dynamics}.
\newblock arXiv:1403, March 2014.

\bibitem{VanKampen1992}
N~G {Van Kampen}.
\newblock {\em {Stochastic processes in physics and chemistry}}, volume~11.
\newblock North- Holland personal library, 1992.

\bibitem{Shahrezaei2008}
Vahid Shahrezaei and Peter~S Swain.
\newblock {Analytical distributions for stochastic gene expression.}
\newblock {\em Proceedings of the National Academy of Sciences of the United
  States of America}, 105(45):17256--61, November 2008.

\bibitem{Grima2012}
R~Grima, D~R Schmidt, and T~J Newman.
\newblock {Steady-state fluctuations of a genetic feedback loop: an exact
  solution.}
\newblock {\em The Journal of chemical physics}, 137(3):035104, July 2012.

\bibitem{Houchmandzadeh2010}
B~Houchmandzadeh and M~Vallade.
\newblock {Alternative to the diffusion equation in population genetics.}
\newblock {\em Phys Rev E Stat Nonlin Soft Matter Phys}, 82(5 Pt 1):51913,
  2010.

\end{thebibliography}

\end{document}